\documentclass{article}
\usepackage{spconf,amsmath,graphicx}
\usepackage{booktabs}
\usepackage{subfigure}
\usepackage[colorlinks,
linkcolor=black,
citecolor=black,
anchorcolor=black,
urlcolor=black]{hyperref}
\title{Claw U-Net: A unet-based NETWORK WITH deep FEATURE CONCATENATION FOR scleral blood vessel segmentation}
\name{Chang Yao$^1$, Jingyu Tang$^1$, Menghan Hu$^{1,*}$, Yue Wu$^2$, Wenyi Guo$^2$, Qingli Li$^1$, Xiao-Ping Zhang$^3$}
\address{$^1$Shanghai Key Laboratory of Multidimensional Information Processing, East China Normal University \\
$^2$Department of Ophthalmology, Ninth People's Hospital Affiliated to Shanghai Jiao Tong University \\ School of Medicine \\
$^3$Department of Electrical, Computer and Biomedical Engineering, Ryerson University
\thanks{This work is sponsored by the Shanghai Education Development Foundation and Shanghai Municipal Education Commission (No. 19CG27).}\\
\thanks{$^*$Corresponding author: Menghan Hu (mhhu@ce.ecnu.edu.cn)}}
\begin{document}
%
\maketitle
\begin{abstract}
Sturge-Weber syndrome (SWS) is a vascular malformation disease, and it may cause blindness if the patient's condition is severe. Clinical results show that SWS can be divided into two types based on the characteristics of scleral blood vessels. Therefore, how to accurately segment scleral blood vessels has become a significant problem in computer-aided diagnosis. In this research, we propose to continuously upsample the bottom layer's feature maps to preserve image details, and design a novel Claw UNet based on UNet for scleral blood vessel segmentation. Specifically, the residual structure is used to increase the number of network layers in the feature extraction stage to learn deeper features. In the decoding stage, by fusing the features of the encoding, upsampling, and decoding parts, Claw UNet can achieve effective segmentation in the fine-grained regions of scleral blood vessels. To effectively extract small blood vessels, we use the attention mechanism to calculate the attention coefficient of each position in images. Claw UNet outperforms other UNet-based networks on scleral blood vessel image dataset.
\end{abstract}
\begin{keywords}
Sturge-Weber syndrome, scleral blood vessel, medical image segmentation, glaucoma, UNet
\end{keywords}
\section{Introduction}
\label{sec:intro}

\begin{figure}[htbp] \centering 

\includegraphics[height=5.2cm, width=9cm]{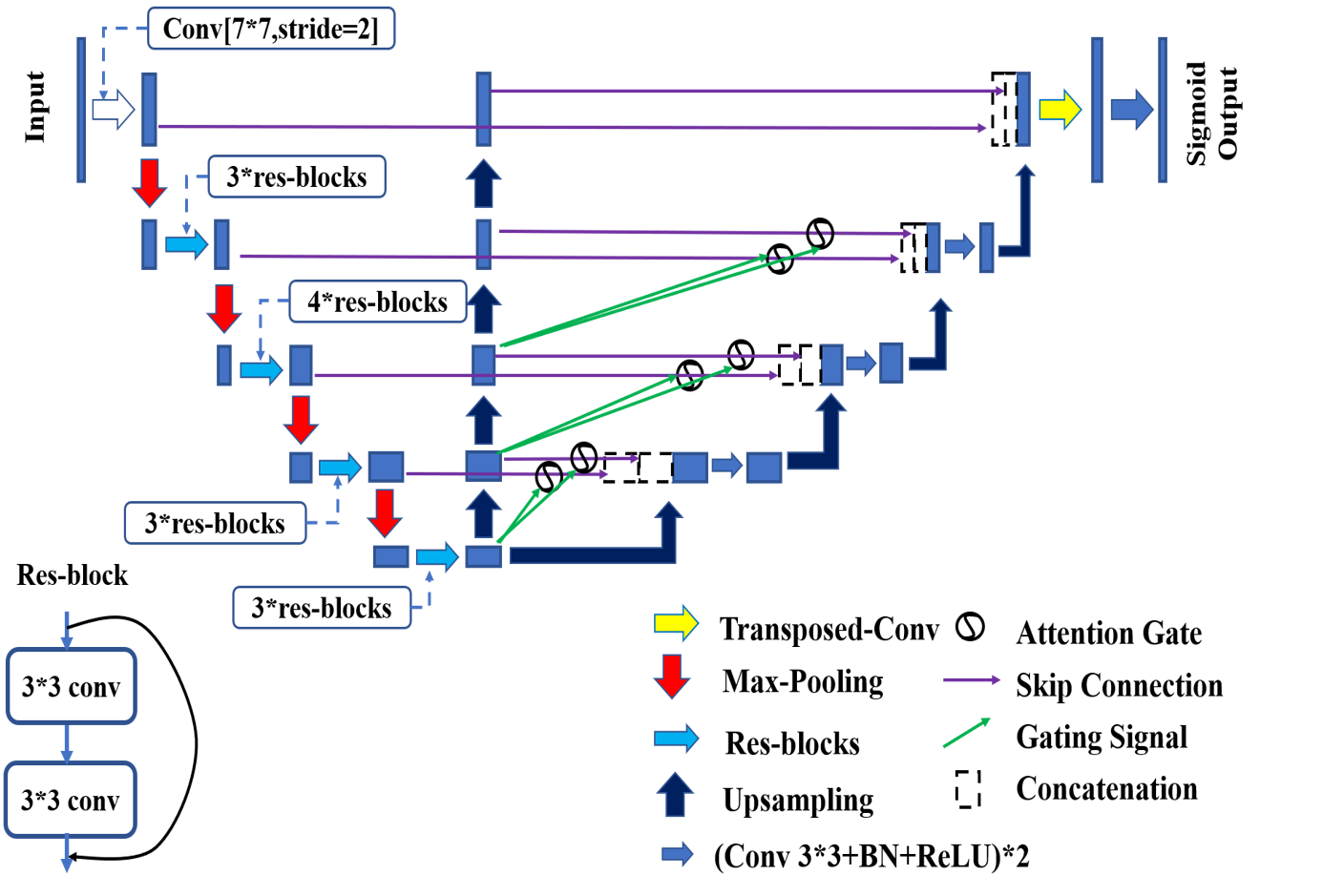} 

\caption{The architecture of Claw UNet.} \label{fig:graph} \end{figure}

Early diagnosis is significant for diseases such as glaucoma, hypertension, and diabetic retinopathy which lead to human vision deterioration \cite{furtado2017segmentation}. Ophthalmologists typically access the clinical condition of retinal blood vessels based on the retinal fundus images, and this is an effective indicator for the diagnosis of various eye diseases \cite{hu2019s}. Sturge-Weber syndrome (SWS) is a vascular malformation disease, and it will cause glaucoma \cite{wu2020episcleral}.  When the symptoms are severe, SWS can cause damage to the skin, brain, and eyes. Glaucoma caused by SWS has two onset peaks viz. onset at birth and onset in adolescence. Due to the seriousness of SWS, it has received great attention from ophthalmologists. Studies have found that the distribution of scleral blood vessels is abnormal for SWS patients. This abnormality may increase the outflow resistance, which in turn leads to glaucoma \cite{mantelli2016ocular} \cite{plateroti2017sturge}. During trabeculotomy, ophthalmologists will divide the patients into two groups based on the degree of blood vessel expansion, whether there is a thick grid-like structure, and whether the blood vessel density of the surgical site is increased. Patients with diffuse vasodilatation and a thick grid-like vascular network have a lower surgical success rate, only 36\% in 2 years, while the other group is as high as nearly 90\% \cite{wu2017early}. The artificial grouping is difficult to promote and requires a high clinical experience. Therefore, there is an urgent need to perform real-time automatic segmentation of scleral blood vessels. It is meaningful for ophthalmologists to take different surgical methods to improve the patient's prognosis and protect the optic nerve to the greatest extent. 

	The existing blood vessel segmentation approaches are mainly designed for fundus images. Unlike fundus images, in scleral vascular images, the vessels to be segmented are denser and have different scales. In addition, it may be not possible to obtain scleral vascular images with high quality during trabeculotomy. With the rapid development of convolutional neural networks (CNNs) \cite{long2015fully}, a variety of end-to-end segmentation models have been developed, such as fully convolutional neural networks (FCNs) \cite{long2015fully}, UNet \cite{ronneberger2015u}, PSPNet \cite{zhao2017pyramid} and  DeepLab \cite{chen2017deeplab}. Among them, UNet shows a good segmentation effect on medical images. The decoder of UNet  provides a high-level semantic feature map, and the encoder provides a low-level detailed feature map. These two phases are combined through skip connections. UNet ++ \cite{zhou2018unet++} improves the strength of these connections by introducing nested and dense skip connections, reducing the semantic difference between encoder and decoder.

	To obtain a good segmentation effect on scleral blood vessel images, we propose a novel UNet-based architecture called Claw UNet by adding skip connections between the deepest feature maps and the decoders. Each decoder part is connected with the upsampling of the bottom layer. By repeatedly using high-level semantic feature maps, the location information in the images can be captured from a complete scale, which helps to accurately segment, especially for the detailed areas of scleral blood vessels. 

	Hence, the main contributions of the current work are in three-fold: 1) we propose a novel UNet-based network Claw UNet, which makes full use of the high-level semantics; 2) we add the attention module and residual structure to make our network concentrate on the boundary segmentation of small blood vessels in the images; 3) we establish the scleral blood vessel image dataset to validate the performance of Claw UNet.

\section{Claw UNet}
\label{sec:method}

Claw UNet combines the encoding and decoding structure of UNet, the nested and dense skip connections of UNet++ to achieve the ideal segmentation effect. Noteworthily, we respectively introduce a residual structure and an attention mechanism in the encoding part and the decoding part to further improve the network performance. 

\subsection{Claw UNet architecture}
\label{ssec:subhead2.1}
The overall structure of Claw UNet is shown in Fig. 1. The architecture can be divided into three parts: the downsampling part obtains the feature maps of input images; the upsampling part restores the segmented images; and the bottom upsampling part retains features to assist segmentation.

The size of input images is set to $512\times512$. The first convolution operation uses kernels with size of $7\times7$, and the stride is set to $2$ to adjust the image size to $256\times256$. Max-pooling operation is used for downsampling to reduce the size of feature maps. To learn deeper information, the residual structure \cite{he2016deep} is applied  for all convolution operations in the encoder. We obtain feature maps $x_{En}^{i, 0}$ with different depths corresponding to each layer, where $i$ represents the layer number of the encoder. The size of feature map of the $i$-th layer is $512/2^{i}$, and the depth is $32\times2^{i}$. The feature maps of the bottom layer are $16\times16$ with the depth of $512$. 

In the decoding part, the feature maps after convolution operation are upsampled. To fuse the extracted features, the decoder $x_{De}^{i, 2}$ is combined with the corresponding encoder feature map $x_{En}^{i, 0}$. The core of our Claw UNet is to send the $x_{Up}^{i, 1}$ that is sampled $4-i$ times from the bottom feature maps to the decoder. This operation allows us to more fully exploit deeper features to maintain more image details. The skip connection can be expressed by the following formula.

when $i=1, \cdots, N-1$,
\begin{small}
\begin{equation}
\mathrm{x}_{\mathrm{De}}^{\mathrm{i} 2}=\left[\mathrm{C}\left(\mathcal{D}\left(X_{\mathrm{En}}^{\mathrm{k}, 0}\right)\right)_{k=1}^{i-1}, \mathrm{C}\left(X_{\mathrm{Up}}^{\mathrm{i}, 1}\right), \mathrm{C}\left(U\left(X_{\mathrm{De}}^{\mathrm{k}, 2}\right)\right)_{k=i+1}^{N}\right]
\end{equation}
\end{small}

when $i = N$,
\begin{small}
\begin{equation}
\mathrm{x}_{\mathrm{De}}^{\mathrm{i}, 2}=\mathrm{x}_{\mathrm{En}}^{\mathrm{i}, 2}
\end{equation}
\end{small}

where $i$ represents the downsampling layer along with the encoder, $N$ represents the total layer of the encoder, $x_{Dn}^{i,2}$ represents the decoding part of the $i-th$ layer. Let function $C(\cdot)$  denotes a convolution operation,  $\mathcal{D}(\cdot)$ and $U(\cdot)$ denote upsampling and downsampling operation separately,  $[\cdot]$ denotes connecting together.

The attention mechanism is introduced between short connections. $x_{En}^{i, 0}$ and $x_{Up}^{i, 2}$ are combined with $x_{De}^{i, 2}$ respectively to calculate the attention coefficient. This operation makes small areas easier to get more attention during segmentation. When the upsampling operation is completed, the deconvolution operation restores the image size from $256\times256$ to the initial size of $512\times512$. Finally, the Sigmoid function makes the output a binary image.

\subsection{Feature learning and fusion}
\label{ssec:subhead2.2}

\begin{figure}[tbp] \centering 

\includegraphics[height=3.5cm, width=9cm]{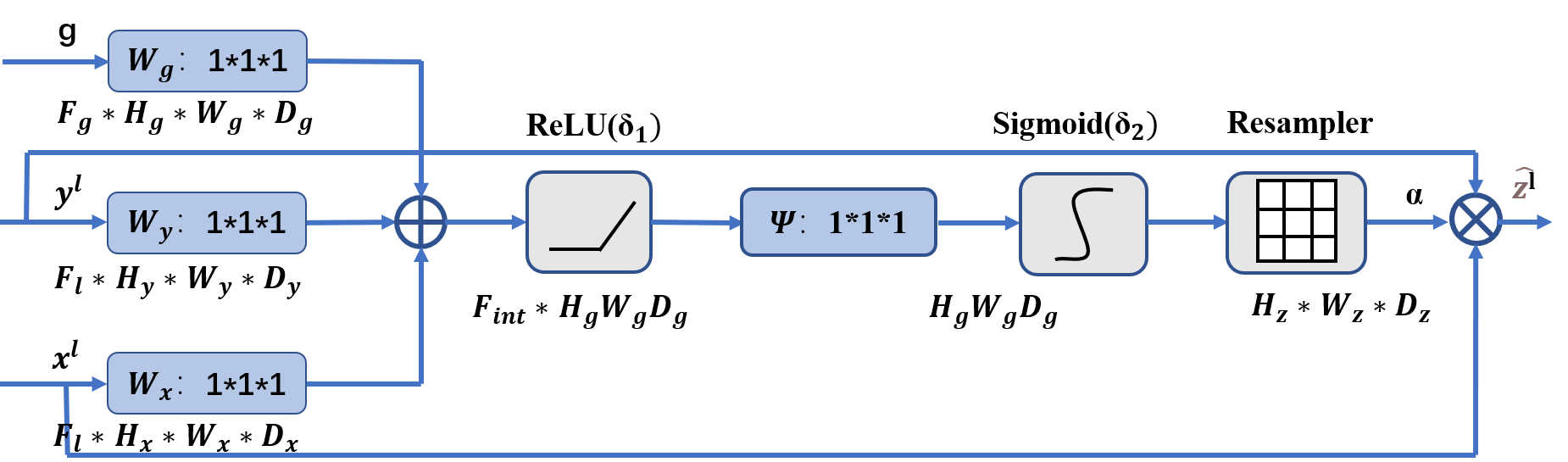} 

\caption{Specific structure inside the attention block. A deep upsampling part is added as the input of the third attention.} \label{fig:graph} \end{figure} 

\begin{figure*}[htbp]
\centering

\includegraphics[width=1\textwidth,height=0.2\textheight]{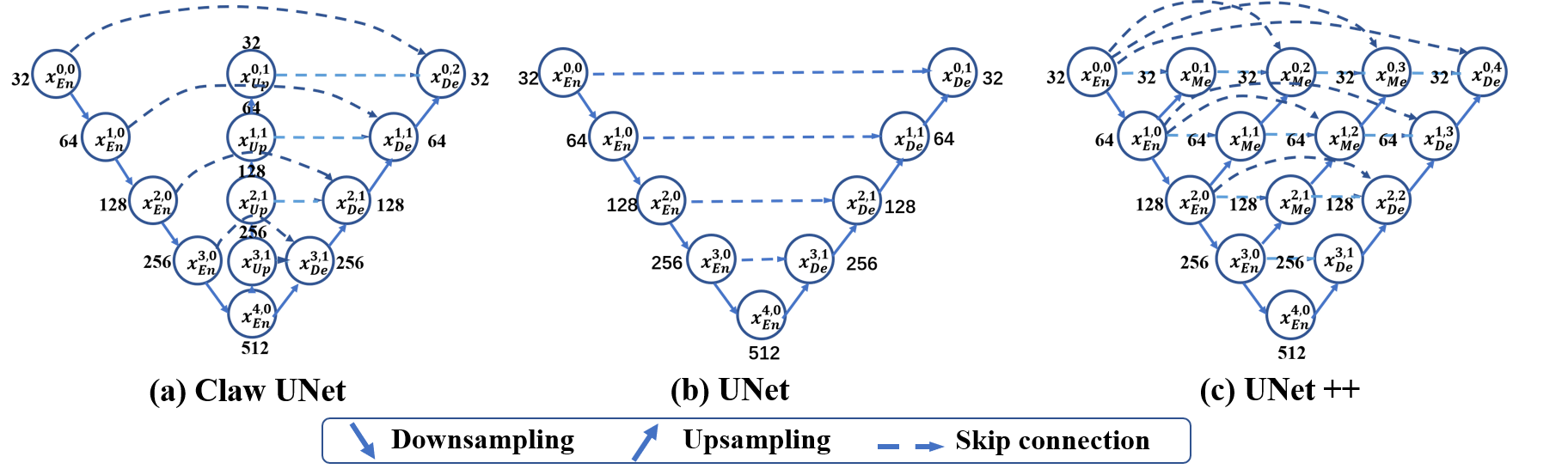}

\caption{Comparison of (a) Claw UNet, (b) UNet, and (c) UNet++.}\label{fig:1}

\end{figure*}

For scleral blood vessel segmentation, many details are not easy to be learned by the network, and therefore we make each encoding part have more convolution operations and deeper network structure. Considering the limited number of actual clinical images, to eliminate the problem of gradient disappearance and explosion in the deep architecture \cite{he2016deep}, we modify the fundamental encoding block to the structure of the residual module, and take Resnet-34 as the backbone of the downsampling part. Each convolution operation uses a $3\times3$ convolution kernel. The residual part is connected by inserting shortcuts. It is worth noting that when inserting connections in the same dimension, no other operations are required. When inserting connections between the layers with different dimensions, additional zero entries are added to achieve dimension matching. 

By integrating the decoding, upsampling and encoding parts, the segmentation of small areas may get better performance. The attention mechanism proposed in \cite{oktay2018attention} uses the features of the next level to supervise the features of the upper level, and optimizes the segmentation by reducing the activation value of the background to achieve end-to-end output.

As shown in Fig. 2, the feature maps in the decoder and the feature maps in the corresponding encoder are sent to the module together. The attention coefficient is more targeted to the local small area, which helps improve performance \cite{li2020arnet}. To maintain more details in images, we also send the feature maps from the bottom upsampling part into the attention block together, and therefore the generated coefficients pay more attention to the deep features.

In the attention block shown in Fig. 2, $g$, $x^{l}$, and $y^{l}$ represent the feature matrix of the decoding part, the encoding part, and the bottom upsampling part resspectively. Following, $x^{l}$ and $y^{l}$ are multiplied by a certain coefficient to achieve attention and then concatenate with $g$. The obtained feature maps are entered into the next layer of the decoding part. The resampler here resamples the feature map to the original size of $x^{l}$ and $y^{l}$. The calculation of the attention coefficient can be expressed by the following formula:

\begin{equation}
\begin{aligned}
q_{a t t}^{l}=& \psi^{T}\left(\sigma_{1}\left(W_{x}^{T} x_{i}^{l}+W_{y}^{T} y_{i}^{l}+W_{g}^{T} g_{i}+b_{g}\right)\right)+b_{\psi} \\
& \alpha_{i}^{l}=\sigma_{2}\left(q_{a t t}^{l}\left(x_{i}^{l}, y_{i}^{l}, g_{i} ; \Theta_{a t t}\right)\right)
\end{aligned}
\end{equation}

where $g$, $y^{l}$, and $x^{l}$ are respectively multiplied by the weight matrix. The weight matrix can be learned through backpropagation to obtain the importance of each element of $g$, $y^{l}$, and $x^{l}$. That is to say, the purpose of introducing attention is to learn the importance of each element and target \cite{oktay2018attention}.

\subsection{Comparison with UNet and UNet++}
\label{ssec:subhead2.3}

According to Fig. 3, we compare the similarities and differences between the proposed Claw UNet, UNet, and UNet++. Fig. 3(b) shows that UNet uses an ordinary skip connection. This decoding process uses the information in the corresponding encoder, and thus many details are often overlooked. Fig. 3(c) shows that UNet++ uses nested and dense skip connections, and the redesigned skip connections aim to reduce the semantic gap between the feature maps of the encoder and decoder \cite{wang2020improved}. Both of the above structures are short of exploring image information on a complete scale. Fig. 3(a) shows that Claw UNet upsamples the feature maps of the bottom layer multiple times and has skip connections among the encoder, the decoder, and the feature maps upsampled from the bottom layer. We believe that when the feature maps from the decoder, the encoder, and the deepest layer are semantically similar, the optimizer will process the details in the image more effectively.

\section{Experiment and results}
\label{sec3}

\begin{table*}[htbp]
\centering 
\caption{\label{tab:test}Performance  comparison  of Claw UNet and other UNet-based networks on scleral blood vessel dataset. The best performer is highlighted in bold.}
 \scalebox{1}[1.6]{
\resizebox{\textwidth}{!}{
 \begin{tabular}{lcccccccccc} 
  \toprule 
  Model & UNet & UNet++ & ResUNet & Channel-UNet & Attention-UNet & R2UNet & ResUNet + attention & UNet++ + attention & Claw UNet & ResClaw Unet + attention  \\ 
  \midrule 
 MIoU (\%) & 79.50 & 79.53 & 79.94 & 78.94 & 79.19 & 75.00 & 78.38 & 79.57 & 78.90 & {\textbf {80.57}} \\ 
 Aver\_hd & 12.83 & 13.01 & 12.22 & 12.54 & 12.54 & 13.59 & 12.89 & 12.25 & 12.66 & {\textbf {11.99}} \\ 
 Dice (\%) & 87.95 & 87.99 & 88.14 & 87.80 & 87.81 & 85.15 & 87.28 & 87.92 & 87.66 & {\textbf {88.58}} \\ 
 
  \bottomrule 
 \end{tabular}}}
 
\end{table*}

\subsection{Experimental Protocol}
\label{ssec:3.1}

\subsubsection{Dataset}
\label{sssec:3.1.1}

We establish the scleral blood vessel image dataset, and the images are taken from the actual surgery. Because it is more difficult to take images during operation, there are certain differences in the size, resolution, and perception field of the captured images. After discussing with experienced ophthalmologists, we intercepted specific blood vessel parts from the images. Ophthalmologists often judge this type of glaucoma caused by SWS based on the blood vessels in these areas. The dataset contains $51$ images of scleral blood vessels taken from 51 different patients and covers two types of such diseases. To facilitate the subsequent network training, we set the size of all images to $512 \times 512$. The masks of the dataset are manually labeled by ophthalmologists from Department of Ophthalmology, Ninth People's Hospital Affiliated to Shanghai Jiao Tong University School of Medicine.

\subsubsection{Evaluation}
\label{sssec:3.1.2}

We divide the dataset for training and testing at a ratio of 4:1 and use binary cross-entropy as the loss function for optimization. For image segmentation problems, Dice is more sensitive to the internal filling of the mask, and Hausdorff distance (Hd) is more sensitive to the segmented boundaries. We finally adopt intersection over union (IoU), Hd, and Dice as indicators to evaluate network performance. For a fair comparison, the parameters of all experiments are set to the same situation. We implement our model on NVIDIA GeForce RTX 2080 Ti using the PyTorch framework.

\subsection{Comparison with other UNet-based Models}
\label{ssec:3.2}

We choose a variety of derivative network structures based on UNet to compare with Claw UNet, including UNet \cite{ronneberger2015u}, UNet++ \cite{zhou2018unet++}, Resnet34-UNet\cite{xiao2018weighted}, Channel-UNet \cite{chen2019channel},  Attention-UNet \cite{oktay2018attention}, R2UNet \cite{alom2018recurrent}. To illustrate the role of residual block and attention block, we add such modules to the comparative network structures. Table 1 shows the comparison results of various networks on the test set.

It can be clearly seen that Claw UNet has achieved good results after combining the residual part and the attention mechanism with MIoU, Dice, and Aver\_hd of $80.57\%$, $88.58\%$, and $11.99$, respectively. MIoUs of other network structures are below $80\%$. To illustrate the necessity of adding residual and attention modules, we test the same network without adding such mechanisms, and MIoU drops by approximately $1.6\%$. At the same time, we add the attention part in UNet++ and Resnet34-UNet, and find that the results are lower than the original networks, showing that the performance improvement not only depends on the increase of modules but also is greatly related to the network structure.

Fig. 4 shows a detailed comparison of the segmentation results of our network and ResUNet. In Fig. 4, Claw UNet preserves more small vascular areas, indicating that Claw UNet is excellent in keeping image details. Small blood vessels are important for ophthalmologists to diagnose disease, and therefore, the sensitivity of Claw UNet to small blood vessels is necessary for computer-aided diagnostic systems.

\begin{figure}[htbp]
\centering
\subfigure[]{
\includegraphics[height=2.8cm, width=3.5cm]{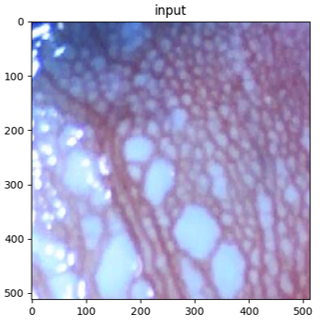}
}
\quad
\subfigure[]{
\includegraphics[height=2.8cm, width=3.5cm]{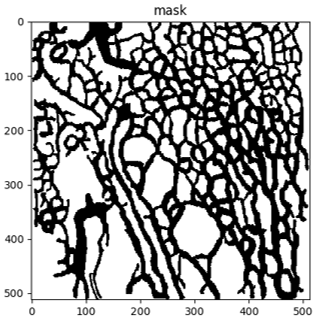}
}
\quad
\subfigure[]{
\includegraphics[height=2.8cm, width=3.5cm]{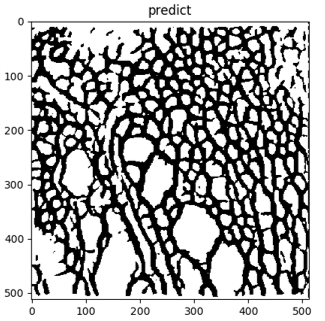}
}
\quad
\subfigure[]{
\includegraphics[height=2.8cm, width=3.5cm]{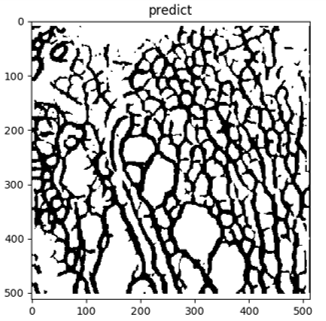}
}
\caption{Visual comparison of segmentation performance between Res-UNet and Claw UNet: a) original image, b) mask,
c) Claw UNet's result, and d) Res-UNet's result.}
\end{figure}

\section{Conclusion}
\label{sec:conclusion}
In this paper, we propose a novel UNet-based network called Claw UNet for the segmentation of scleral blood vessel images. The upsampling parts from the bottom layer provide rich detailed information to improve small blood vessels segmentation. To achieve more precise segmentation, residual structure and attention mechanism are introduced into Claw UNet. The residual structure can effectively extract deep information, the deepest level features can retain more details of the original image, and the attention mechanism can produce accurate boundary perception. In addition, we establish the scleral blood vessel image dataset to validate the performance of Claw UNet. The experimental results show that Claw UNet is more effective and superior to previous work in detail segmentation of scleral blood vessel.

\small
\bibliographystyle{IEEEbib}
\bibliography{strings,refs}

\end{document}